\documentclass[prl,twocolumn,superscriptaddress]{revtex4}
\usepackage{graphicx}
\usepackage{amssymb}
\usepackage{amsmath}
\usepackage{bm}
\usepackage{color}
\pdfoutput=1

\begin{document}

\title{Talbot effect for exciton-polaritons}

\author{T. Gao}
\affiliation{Research School of Physics and Engineering, The Australian National University, Canberra ACT 2601, Australia}

\author{E. Estrecho}
\affiliation{Research School of Physics and Engineering, The Australian National University, Canberra ACT 2601, Australia}

\author{G. Li}
\affiliation{Research School of Physics and Engineering, The Australian National University, Canberra ACT 2601, Australia}

\author{O. A. Egorov}
\affiliation{Institute of Condensed Matter Theory and Solid State Optics, Abbe
Center of Photonics, Friedrich-Schiller-Universit\"at Jena, Max-Wien-Platz 1, 07743 Jena, Germany}

\author{X. Ma}
\affiliation{Department of Physics, Center for Optoelectronics and Photonics Paderborn (CeOPP), Universit\"at Paderborn, Warburger Strasse 100, 33098 Paderborn, Germany.}

\author{K. Winkler}
\affiliation{Technische Physik, Wilhelm-Conrad-R\"ontgen-Research Center for Complex
Material Systems, Universit\"at W\"urzburg, Am Hubland, D-97074 W\"urzburg,
Germany}

\author{M. Kamp}
\affiliation{Technische Physik, Wilhelm-Conrad-R\"ontgen-Research Center for Complex
Material Systems, Universit\"at W\"urzburg, Am Hubland, D-97074 W\"urzburg,
Germany}

\author{C. Schneider}
\affiliation{Technische Physik, Wilhelm-Conrad-R\"ontgen-Research Center for Complex
Material Systems, Universit\"at W\"urzburg, Am Hubland, D-97074 W\"urzburg,
Germany}

\author{S. H\"ofling}
\affiliation{Technische Physik, Wilhelm-Conrad-R\"ontgen-Research Center for Complex
Material Systems, Universit\"at W\"urzburg, Am Hubland, D-97074 W\"urzburg,
Germany}
\affiliation{SUPA, School of Physics and Astronomy, University of St Andrews, St Andrews
KY16 9SS, United Kingdom}

\author{A. G. Truscott}
\affiliation{Research School of Physics and Engineering, The Australian National University, Canberra ACT 2601, Australia}

\author{E. A. Ostrovskaya}
\affiliation{Research School of Physics and Engineering, The Australian National University, Canberra ACT 2601, Australia}


\begin{abstract}
We demonstrate, experimentally and theoretically, a Talbot effect for hybrid light-matter waves --- exciton-polariton condensate formed in a semiconductor microcavity with embedded quantum wells. The characteristic `Talbot carpet' is produced by loading the exciton-polariton condensate into a microstructured one dimensional periodic array of mesa traps, which creates an array of sources for coherent polariton flow in the plane of the quantum wells. The spatial distribution of the Talbot fringes outside the mesas mimics the near-field diffraction of a monochromatic wave on a periodic amplitude and phase grating with the grating period comparable to the wavelength. Despite the lossy nature of the polariton system, the Talbot pattern persists for distances exceeding the size of the mesas by an order of magnitude. 
\end{abstract}
\maketitle

{\em Introduction.---} Talbot effect is a manifestation of {\em near-field} (Fresnel) diffraction of a coherent plane wave incident on a periodic grating, which results in a nontrivial 2D pattern of fringes often referred to as a `Talbot carpet'. According to the Huygens-Fresnel principle, it is interpreted as interference of coherent spherical waves originating from the apertures of the grating. Nearly two centuries after the discovery of the optical Talbot effect \cite{Talbot1836}, it continues to be re-discovered and re-examined in the context of matter and optical waves of various physical nature and spatial scales. Apart from the linear and nonlinear optics \cite{Wen_review_2013}, the Talbot effect has been observed with atoms \cite{Talbot_Lau,Talbot_atoms,Talbot_atoms_mw}, electrons \cite{Talbot_electron_NJP}, X-rays \cite{Talbot_Xray,Talbot_Xray_03}, single photons \cite{Talbot_photons}, and surface plasmon-polaritons (SPPs) \cite{Talbot_SPP_Dennis,Talbot_SPP_optex,Talbot_SPP_nano}. The visually stunning effect is not purely of aesthetic value. Talbot interference has a deep connection with number theory and theory of quantum revivals \cite{Berry96,Berry}, and serves a range of practical purposes. Indeed, periodic self-imaging of the source resulting from the Talbot effect \cite{Talbot_Rayleigh} gives rise to grating-based imaging techniques \cite{Talbot_Xray,Talbot_Xray_03,Talbot_X ray imaing}. Various applications in metrology, atomic lithography, and optical manipulation have also been suggested \cite{Wen_review_2013}.

The common prerequisite for the observation of the Talbot effect is a periodic arrangement of sources of coherent spherical waves that can propagate in the direction perpendicular to the direction of the array. In optics, this is naturally achieved by diffraction of an incident light on an array of apertures \cite{Wen_review_2013}. However, this effect can also be reproduced by other means. For example, in SPP physics Talbot interference of waves from periodically arranged sources rather than diffraction of a plane wave on a periodic grating has been observed \cite{Talbot_SPP_nano}.

\begin{figure}
\includegraphics[width=8.5 cm]{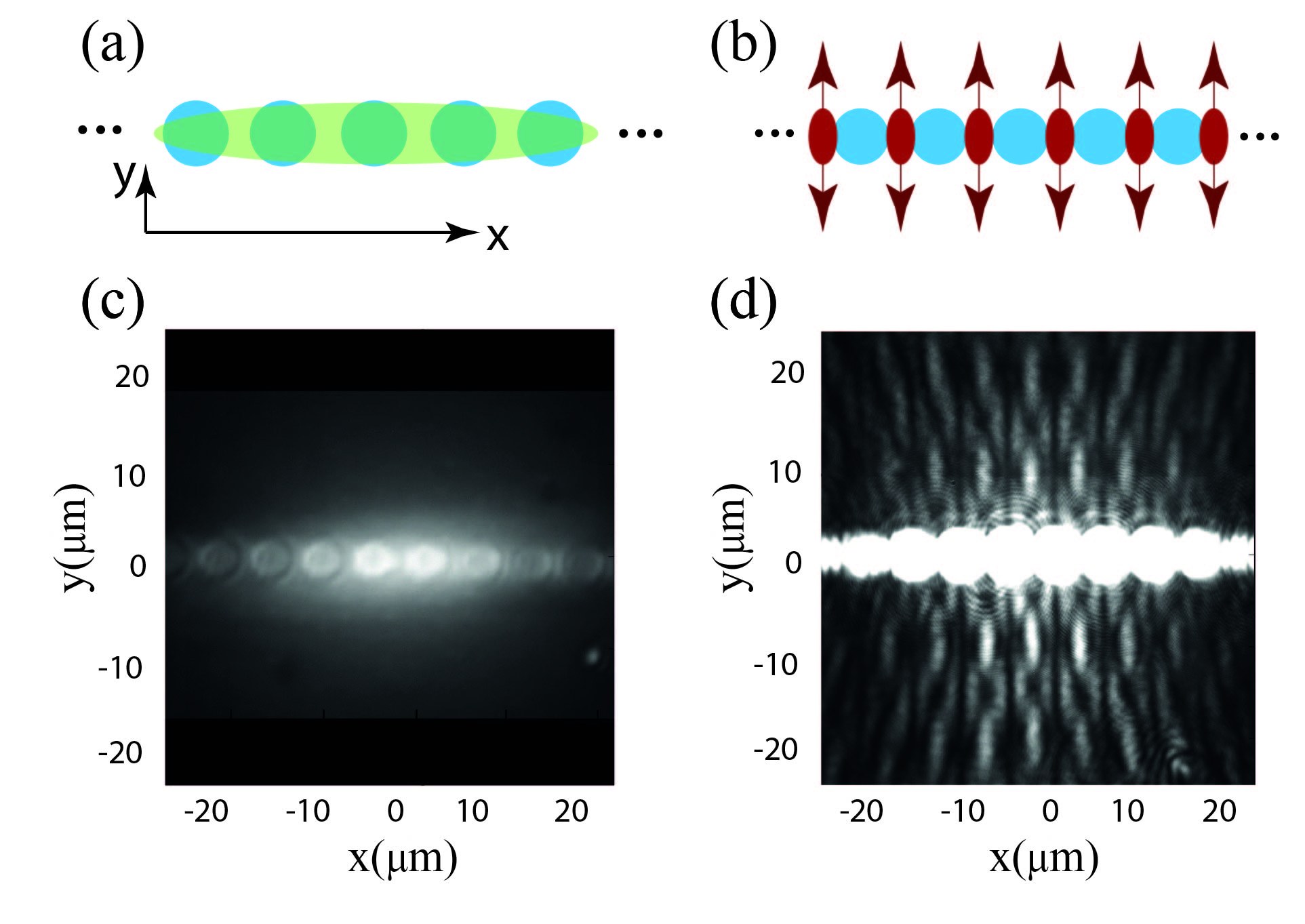}
\caption{Schematics of (a) an elliptical optical pump illuminating a 1D array of mesa traps in a microcavity and (b) polariton flows (arrows) responsible for the Talbot effect in (d). Real space images of the exciton-polariton emission (c) at low excitation power, below the condensation threshold and (d) at high excitation power, above the condensation threshold. The Talbot effect is visible in (d).} 
\label{real_space}
\end{figure}

Recently, experimental investigations of macroscopic coherent quantum states (bosonic condensates) of microcavity exciton-polaritons \cite{BEC06,CiutiREV13,YamamotoREV14} in one-dimensional (1D) periodic potentials has received a lot of attention \cite{stanford1d,gap_state,acoustic_tight,arXiv_15,frustrated}. Exciton-polaritons are quasiparticles arising due to hybridisation of an electron-hole pair (exciton) and photon in a strong light-matter coupling regime. To achieve the strong coupling, the exciton is spatially confined in a 2D quantum well grown in an optically pumped semiconductor microcavity, the latter ensuring resonance with a long-lived photon mode. Periodic potentials for polaritons can be formed by surface metal deposition on the semiconductor microcavity \cite{stanford1d,stanford2d,stanfordkagome}, acousting wave modulation \cite{acoustic,acoustic_tight}, deep etching of micropillars \cite{gap_state,frustrated} or microstructuring of buried mesa traps \cite{NJP_15,arXiv_15} (see \cite{trapping_review} and references therein). These potentials act as a solid-state superlattice: when a microcavity is optically pumped, the exciton-polaritons populate the energy bands of the resulting periodic potentials. Propagation of coherent polariton waves in the plane of a quantum well, away of the periodic array, has never been investigated in detail because it is difficult, if not impossible, to avoid trapping of polaritons in stationary (extended or localised) states of the 1D array \cite{gap_state,acoustic_tight}.

In this work, we employ an exciton-polariton condensate in a 1D buried mesa array of polariton traps [see schematics in Fig. \ref{real_space}(a,b)] to observe the Talbot interference patterns with coherent hybrid light-matter waves. The Talbot effect for exciton-polaritons is uniquely enabled by the ability of exciton-polaritons to condense into a non-ground extended (Bloch) state of the 1D array \cite{NJP_15,arXiv_15}, as well as by the nature of the mesa traps, which are embedded into the microcavity \cite{NJP_15} rather than forming free-standing pillars on the substrate \cite{gap_state}. When the Bloch mode is characterized by a periodic distribution of polariton density maxima located in the {\em barriers} between mesa traps, the barrier regions act as sources of polariton waves, which are free to propagate in the plane of the quantum well [see Fig. \ref{real_space}(b)]. The periodic array of such sources creates a Talbot carpet shown in Fig. \ref{real_space}(d). Moreover, we demonstrate that this system mimics both amplitude and phase gratings for the light-matter waves.

{\em Experiment.---} The experiment was performed using 1D mesa arrays microstructured in an AlAs/AlGaAs microcavity with GaAs quantum wells, as described in \cite{NJP_15}. Mesas of $3.5$~ $\mu$m diameter are separated centre-to-centre by the distance of $5.5$~$\mu$m, with the effective polariton potential depth of $\sim 5$ meV for each mesa. The exciton-polariton condensate is formed spontaneously by pumping the microcavity with a cw laser injecting free carriers well above the polariton energy. The pump beam has a FWHM dimension of $2.5 \times 36$~$\mu$m, which illuminates approximately $6$ mesas, as shown in Fig.~\ref{real_space}(c). Real and reciprocal space imaging of the  cavity photoluminescence resulting from the polariton decay is used to analyse the spatial density distribution and dispersion of exciton-polaritons.

\begin{figure}
\includegraphics[width=8.5 cm]{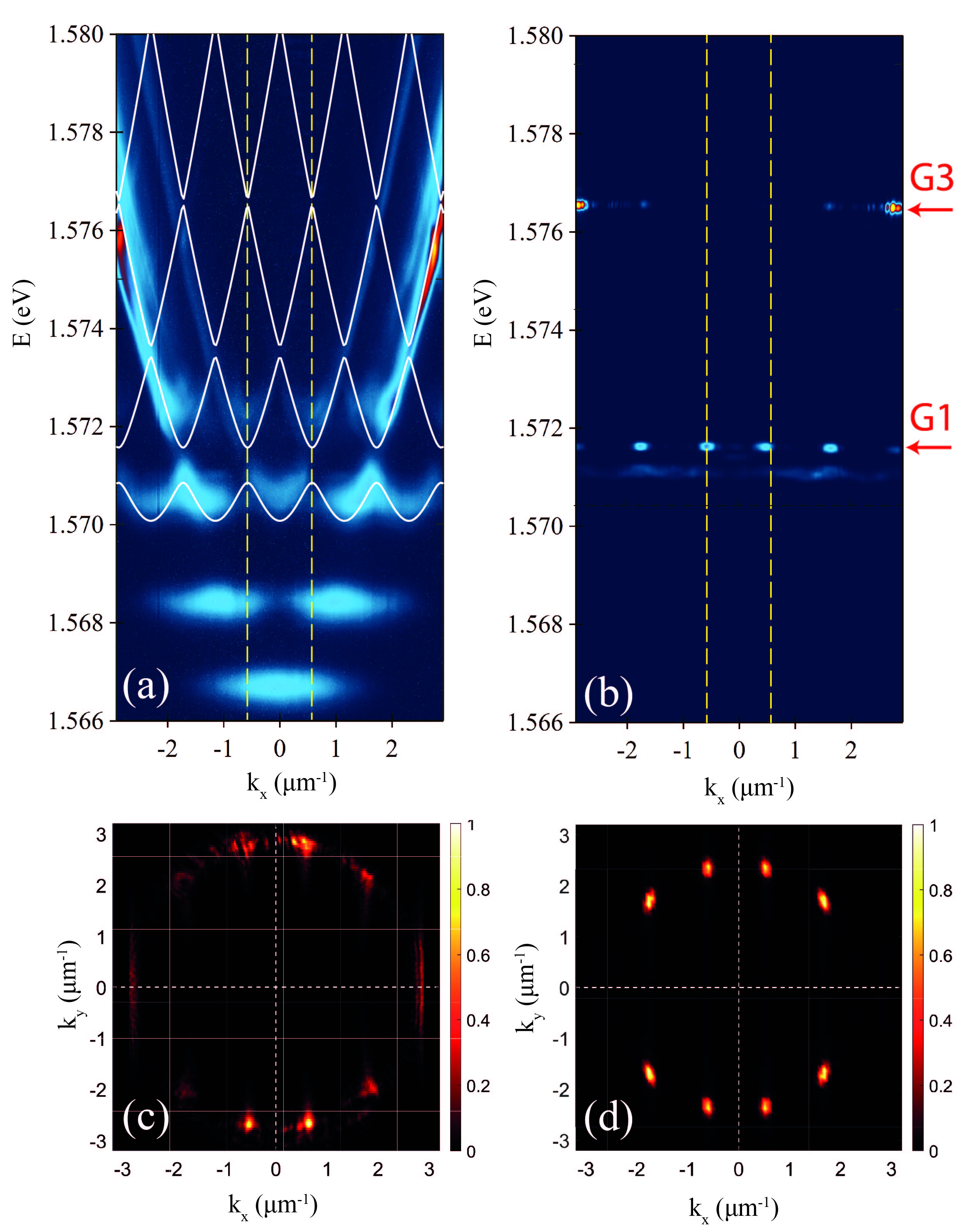}
\caption{(a,b) Dispersion measurement of polariton emission from the mesa traps (a) below and (b) above condensation threshold. Both localised and extended energy states are seen in (a). Condensation in two gap states in the first (G1) and the third (G3) spectral gaps are visible in (b). Dashed yellow lines correspond to $\pm k_B/2$ and mark the first Brillouin zone of the array. The solid lines in (a) show the first four extended Bloch bands calculated from Eq.~(\ref{eq_schrodinger}). (c,d) Energy filtered reciprocal space image of the condensate emission from the G3 state (c) measured in the experiment, and (d) calculated theoretically from the field distribution $f_{T}(\bm{r})$ (see text). } 
\label{dispersion}
\end{figure}

In the regime of low excitation powers, the dispersion (energy vs. in-plane momentum) of exciton-polaritons created outside and in the mesa array are remarkably different. Outside the mesa array, the parabolic dispersion $E({\bm k})$ near the in-plane momentum ${\bm k}=0$ is typical of the lower polariton dispersion branch in a planar microcavity, as seen in the Supplemental Material (SM) \cite{SM}. In contrast, the emission from polaritons located in mesa traps reveals the band-gap energy structure \cite{lattice_PRL} imposed by the periodicity of the trapping potential in the lateral ($x$) direction [Fig.~\ref{dispersion}(a)], as described in \cite{NJP_15,arXiv_15}. Both the discrete energy states in the individual mesas and the characteristic band-gap spectrum of extended Bloch states can be seen in Fig.~\ref{dispersion}(a). The lowest band of Bloch states is formed above the excited energy state in the individual mesas, similarly to the spectra of the deep photonic wires in organic microcavities \cite{wires_2014,wires_2015}. The gaps between the energy bands become  progressively narrower, as can be seen from the spectrum in Fig.~\ref{dispersion}(a). 

At the higher pump powers, exciton-polaritons undergo transition to bosonic condensation. 
The majority of the condensate populates non-ground steady states marked by G1 and G3 arrows in Fig.~\ref{dispersion}(b). These are weakly spatially localised gap states \cite{bec_gap} that, in the narrow gaps, are indistinguishable from Bloch modes modulated by a broad Gaussian envelope. In particular, the highly populated steady state G3 forms in the very narrow third energy gap of the linear spectrum in Fig.~\ref{dispersion}(a). Its density and phase distributions are inherited from the stationary Bloch state at the top of the third spectral band. Unlike propagating polariton waves, the steady states trapped in the array are nearly monochromatic.

The reciprocal ($k$) space image of the emission intensity also undergoes dramatic changes above the exciton-polariton condensation threshold. At the low power, the $k$-space image exhibits a ring-shape distribution due to the flow of the untrapped polaritons in all directions. Once the pump power exceeds the condensation threshold, polaritons condense predominantly in the G3 mode determined  by the extended Bloch state of the mesa array. This state is characterised by two peaks at the edges of the third Brillouin zone $k_x = \pm 2.85 \mu m^{-1}$ visible in Fig.~\ref{dispersion}(b) and the corresponding maxima in the $k$-space emission pattern Fig.~\ref{dispersion}(c). Moreover, Fig.~\ref{dispersion}(c) demonstrates discrete distribution of the polariton emission intensity in the $k$-space which suggests additional periodicity in the transverse ($y$) direction. This is the direct consequence of the Talbot effect, which is revealed in the spatial distribution of the polariton density.

\begin{figure}
\includegraphics[width=8.5 cm]{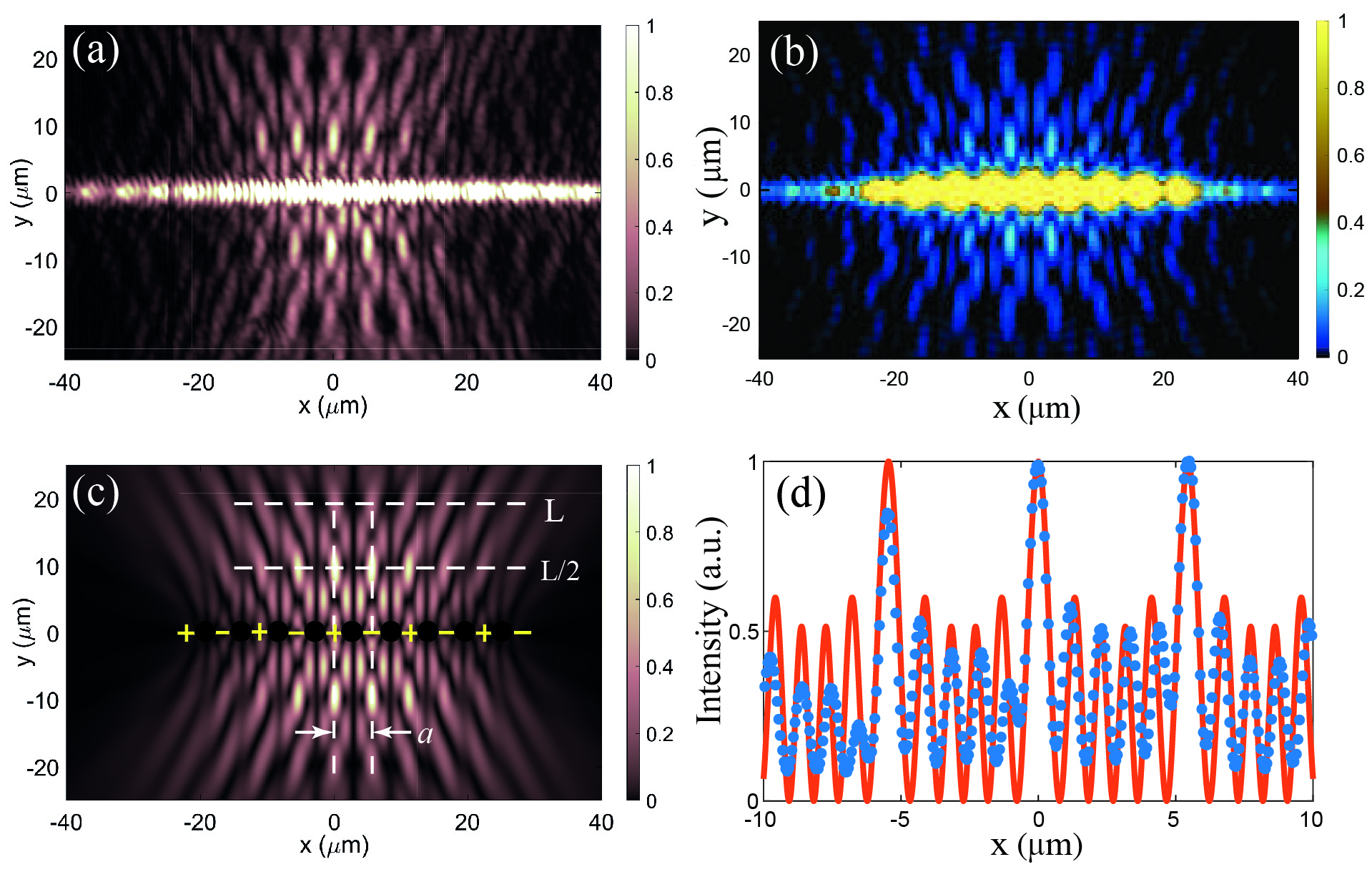}
\caption{(a) Experimental spatial distribution of the Bloch mode intensity and the Talbot pattern [energy filtered at the G3 position in Fig.~\ref{dispersion}(b)]. (b,c) Talbot pattern calculated (b) numerically using the full nonlinear 2D mean-field model and (c) theoretically using the Huygens-Fresnel superposition. The "+" and "-" signs in (c) indicate relative $\pi$ phase difference between the adjacent sources of the polariton flow located between the mesas (black circles), and $L$ marks the Talbot length; (d) Comparison between $|\phi_B(x)|^2$ calculated using Eq.~(\ref{eq_schrodinger}) (solid line) and the experimental real-space profile (circles) taken along the line $y=0$ in (a). } 
\label{pattern_compare}
\end{figure}

In the  G1 state, the majority of the polariton density is contained within the individual mesas. In contrast, the real space density distribution of the G3 state along the $y=0$ line shown in Figs~\ref{pattern_compare}(a,d) displays larger polariton density in the potential barrier regions between the individual mesas. Polaritons in the barrier regions are free to propagate in the plane of the quantum well, thus generating coherent flow of polaritons in the transverse direction. Real space image of the polariton flow outside the mesa array [Figs~\ref{pattern_compare}(a)]  shows the interference structure consistent with the linear Talbot effect, with the spatial periodicity both in the lateral and transverse directions. 

With the growing pumping power above the condensation threshold, the contrast of the Talbot fringes becomes enhanced as discussed in SM \cite{SM}.




{\em Theory.---} The full dynamics of exciton-polariton condensation in the one-dimensional mesa array for moderate pump powers above threshold can be reliably reproduced by the two-dimensional mean-field dynamical model taking into account energy relaxation due to quantum and thermal fluctuations in the system \cite{Wouters2008}. The detailed description of the model, which describes transition between different energy states occupied by the condensate in the mesa traps can be found in \cite{arXiv_15}. It consists of the open-dissipative Gross-Pitaevskii equation for the condensate wavefunction incorporating stochastic fluctuations and coupled to the rate equation for the excitonic reservoir created by the off-resonant cw pump \cite{Wouters2008,arXiv_15}.  Numerical modelling with the parameters corresponding to our experiment reproduces condensation into a non-ground steady state state with the real space density distribution shown in Fig.~\ref{pattern_compare}(b). It reveals the Talbot pattern in qualitative agreement with the experiment. In what follows, we present a simple, intuitive theory of this effect based on its analogy with the linear near-field diffraction.  


In the low density regime, i.e., for the excitation powers below the condensation threshold, the exciton-polaritons occupy the band-gap ladder of single-particle energy states in a periodic potential (see, e.g., \cite{NJP_15} for details of the potential characterisation). The energy band structure $E_n({\bm k})$, where $n$ is the band index, can be calculated directly by solving the stationary single-particle Schr\"odinger equation for the macroscopic wavefunction of the polariton condensate $\psi(x,y)$ in a periodic in-plane potential. Due to the 1D nature of the lateral periodicity, the dimensionality reduction can be performed, and eigenvalues approximated by the spectrum of the factorised eigenstates $\psi(x,y)=\chi(y)\phi_k(x)\exp(-ikx)$, where $\chi(y)$ is a transverse mode of an individual mesa trap, and $\phi_{k} (x)=\phi_{k} (x+a)$ are the extended polariton Bloch states in an {\em effective} 1D  potential $V(x)=V(x+a)$ with the in-plane momentum $k\equiv k_x$. The 1D Bloch states obey the following equation:
\begin{equation} \label{eq_schrodinger}
\left[\frac{\hbar^2}{2m_{p}} \left(k-i\nabla_x\right )^2+V(x)\right] \phi_{n,k} (x)=E_n (k) \phi_{n,k} (x).
\end{equation}
Here $m_p \approx 4.45\times 10^{-5} m_e$ is the effective polariton mass in the planar region, and $m_e$ is the free electron mass. We approximate the 1D periodic potential created by the mesas by the anharmonic analytical function $V(x)=V_0\left [F(x)-1 \right ]$, where:
\begin{equation}
F(x)= \frac{(1+s)^2 \left[ 1+\cos(k_B x)    \right] }{2\left[ 1+s^2+2s\cos(k_B x) \right ]}, \nonumber
\end{equation}
and $k_B=2\pi/a$ is the size of the Brillouin zone of the periodic potential. The shapes of the potential for varying degrees of anharmonicity are described in \cite{anharmonic}.

The band-gap spectrum $E_n(k_x)$ calculated using Eq.~(\ref{eq_schrodinger}) for the anharmonicity parameter $s=-0.2$ and the potential depth $V_0=5.2$ meV, assuming the transverse ground state of the mesa $\chi(y)=\chi_0(y)$, demonstrates good agreement with the experimentally measured spectrum, as seen in Fig.~\ref{dispersion}(a). We note, however, that the multitude of populated energy bands visible in Fig.~\ref{dispersion}(a) includes extended states formed by hybridisation of the higher-order two-dimensional modes of the individual mesa traps. One of such bands, weakly populated by low-density polaritons, is seen in Fig.~\ref{dispersion}(b) below the state G1.

The lateral density distribution for the polariton Bloch mode giving raise to the G3 state emission in Fig.~\ref{dispersion}(b) is well reproduced by our model. The calculated density of the Bloch state $\phi_B(x) \equiv \phi_{3,k_B/2}(x)$ and the corresponding experimental intensity of the polariton emission along the $x$ axis are shown in Fig.~\ref{pattern_compare}(d). The highest peaks of $|\phi_B(x)|$ are located in the local maxima of $V(x)$ which correspond to the barrier regions of the mesa array. This Bloch state belongs to the top edge of the third energy band ($n=3$) and therefore is a '$\pi$-state' with a {\em staggered phase}, i.e. its adjacent density peaks have $\pi$ phase difference. This phase difference is inherited by the transverse flow of polaritons, which in turn leads to some characteristic features in the Talbot pattern. Conversely, from the {\em intensity} of the polariton emission in the Talbot pattern one can reliably infer the {\em phase} distribution of the condensate wavefunction $\phi_B(x)$ in the mesa array. 

The Talbot pattern can be reproduced by applying the linear Huygens-Fresnel principle to the polariton flow, thus assuming that polariton waves propagating away from the mesa array are of sufficiently low particle density, so that the nonlinearity can be ignored. As indicated by the lateral period of the Talbot pattern, only the highest peaks of $|\phi_B(x)|^2$ act as sources of the coherent polariton flow. According to the Huygens-Fresnel principle, the condensate in the barrier regions can be represented by an array of point sources of decaying radial waves whose field  is given by \cite{Wouters2008}: $f_m\sim(1/\sqrt{r})\exp[ i (k_p r_m+\theta_m)-\kappa r_m]$, where $r_m=\sqrt{| \bm{r} - \bm{r}_m |^2}$ is the relative distance between a point on the plane, $\bm{r}$, and the position of the $m$-th source, $\bm{r}_m$, $k_p=2\pi/\lambda_p$ is the polariton wavevector, $\kappa =\gamma m_p/(2\hbar k_p)$ is determined by the polariton decay rate $\gamma$, and $\theta_m$ is the initial phase inherited from $\phi_B(x)$. The total field of the Talbot carpet is given by the linear superposition of waves from all sources: $f_{T}(\bm{r})=\sum_m f_m$, which, in our case, are located on the line $y=0$.

Figure~\ref{pattern_compare}(c) shows the Talbot pattern reproduced by the linear theory assuming the polariton wavelength $\lambda_p=2.6$~$\mu$m and the polariton lifetime $\gamma^{-1}\sim10$~ps. The real space image is in excellent agreement with the experimental pattern Fig.~\ref{pattern_compare}(a). The dark lines located between bright lobes result from destructive interference between the adjacent sources due to the relative $\pi$ phase difference. Away from the $x$ axis, the pattern distorts and eventually vanishes because of the finite number of sources and polariton decay. Nevertheless, the pattern persists for distances sufficient to re-image the source twice. Finally, the calculated Fourier transform of the Talbot pattern $f_{T} (\bm{r})$ matches the experimental $k$-space signature very well [cf. Figs.~\ref{dispersion}(c) and ~\ref{dispersion}(d)]. The discrepancies between the calculated and experimental images stem from the theoretical assumption of distributed point sources located on $y=0$, whereas in practice each source has a finite transverse extent.

The Talbot length is defined as the transverse distance at which the phase shift of all plane wave sources is equal to $2\pi N$, where $N$ is an integer. At this distance, the original density distribution at $y=0$ is revived. Remarkably, in our case the Talbot pattern demonstrates the complete revival at half the Talbot length $L/2$ [see Figs.~\ref{pattern_compare}(a,c)]. This is because the array of coherent polariton sources created by the Bloch state mimics both amplitude and {\em phase} gratings. At the transverse distance $L/2$, the phase of each source acquires a $\pi$ shift, thus reproducing the staggered phase structure and the polariton density (emission intensity) pattern of the origin. The Talbot length determined from the transverse period of the pattern in the experimental real-space image is  $L\approx 20$ $\mu$m. The wavelength of the coherent polariton wave is comparable to the period of the mesa array ($a=5.5$~$\mu$m), and can be calculated using the non-paraxial correction to the Rayleigh formula \cite{Talbot_Rayleigh,Talbot_SPP_nano}: $L=\lambda_p [1-\sqrt{1-\lambda^2_p /a^2}]^{-1}$, which yields $\lambda_p \approx 2.8$ $\mu$m. This value agrees well both with that assumed in our theoretical calculations above, and with the de Broglie wavelength of the transversely free polaritons with the energy $E_p=5.81$ meV [state G3 in Fig.~\ref{dispersion}(b)]: $\lambda_{dB}=h/\sqrt{2m_pE_p}=2.4$ $\mu$m.

Finally, we note that the Talbot interference of exciton-polaritons is an ubiquitous effect. However, not every higher-order exciton-polariton mode responsible for leaking into the planar regions and generating Talbot patterns may be captured by our simple 1D theory, as discussed in SM \cite{SM}.

%

{\em Conclusions.---} To summarise, we demonstrate the Talbot effect for an exciton-polariton condensate loaded into a one-dimensional array of mesa traps. The Talbot pattern is formed due to condensation of exciton-polaritons into a non-ground energy state which has pronounced density peaks in the barrier regions between mesas. Polaritons at these locations are free to propagate transversely to the mesa array and therefore the barrier regions act as a periodic array of sources of coherent polariton waves. Due to the nontrivial phase of the non-ground state in the array, these sources mimic both {\em amplitude} and {\em phase} gratings for the light-matter waves, which links our observations to the Lohmann effect \cite{Lohmann90}. Moreover, the period of the grating is {\em comparable} to the wavelength of the coherent light-matter waves, which provides opportunity for exploring non-paraxial effects on the microscale. Numerical calculations exploiting a mean-field nonlinear model of polariton condensation, as well as the linear Huygens-Fresnel diffraction theory, allow us to reproduce the spatial distribution and spectral signatures of the Talbot effect. 

Our results represent the first, to the best of our knowledge, observation of the Talbot effect with hybrid light-matter waves. This research opens the avenue for using the current advanced nanofabrication techniques to engineer flow patterns of exciton-polaritons in the plane of the quantum wells embedded in the microcavity. In particular, lensing, beam splitting, and phase-dependent shaping of polariton flows enabled by the Talbot effect could be realised \cite{Talbot_lensing,Talbot_SPP_optex}. The resulting control over in-plane polariton propagation could aid the development of integrated polaritonic devices.

{\em Acknowledgements} This work was supported by the Australian Research Council and the state of Bavaria. O.A.E. acknowledges financial support by the Deutsche Forschungsgemeinschaft (DFG project EG344/2-1). Assistance in the sample fabrication by M. Emmerling, A. Schade and J. Gessler is gratefully acknowledged.


\begin{thebibliography}{99}
\bibitem{Talbot1836} H. F. Talbot, ``Facts relating to optical science, No. IV," Philos. Mag. {\bf 9}, 401 (1836).
\bibitem{Wen_review_2013} J. Wen,Y. Zhang, and M. Xiao, ``The Talbot effect: recent advances in classical optics, nonlinear optics, and quantum optics", Adv. Opt. and Phot. {\bf 5}, 83 (2013).
\bibitem{Talbot_Lau} J. F. Clauser and Sh. Li, ``Talbot-von Lau atom interferometry with cold slow potassium", Phys. Rev. A {\bf 49}, R2213 (1994)
\bibitem{Talbot_atoms} M. S. Chapman, Ch. R. Ekstrom, T. D. Hammond, J. Schmiedmayer, B. E. Tannian, S. Wehinger, and D. E. Pritchard, ``Near-field imaging of atom diffraction gratings: The atomic Talbot effect", Phys. Rev. A {\bf 51}, R14 (1995)
\bibitem{Talbot_atoms_mw} S. Nowak, Ch. Kurtsiefer, T. Pfau, C. David, ``High-order Talbot fringes for atomic matter waves", Opt. Lett. {\bf 22}, 1430 (1997).
\bibitem{Talbot_electron_NJP} B. J. McMorran and A. D. Cronin, ``An electron Talbot interferometer", New J. Phys. {\bf 11}, 033021 (2009)
\bibitem{Talbot_Xray}  C. David, B. N\"ohammer, H. H. Solak, E. Ziegler,  ``Differential x-ray phase contrast imaging using a shearing interferometer". Appl. Phys. Lett. {\bf 81}, 3287 (2002).
 \bibitem{Talbot_Xray_03} A. Momose, Sh. Kawamoto, I. Koyama, Y. Hamaishi, K. Takai, and Y. Suzuki, ``Demonstration of X-Ray Talbot Interferometry", Japan. J. Appl. Phys. {\bf 42}, L866 (2003)
\bibitem{Talbot_photons} X.-B. Song, H. B. Wang, J. Xiong, K. Wang, X. Zhang, K.-H. Luo, and L.-A. Wu, ``Experimental Observation of Quantum Talbot Effects", Phys. Rev. Lett. {\bf 107}, 033902 (2011)

\bibitem{Talbot_SPP_Dennis} M. R. Dennis, N. I. Zheludev, and F. J. Garc\'ia de Abajo, ``The plasmon Talbot effect", Opt. Exp.{\bf 15}, 9692 (2007)
\bibitem{Talbot_lensing} F. M. Huang, Y. Chen, N. I. Zheludev, and F. J. Garc\'ia de Abajo, ``Focusing of light by a nanohole array", Appl. Phys. Lett. {\bf 90}, 091119 (2007)
\bibitem{Talbot_SPP_optex} W. Zhang, Ch. Zhao, J. Wang, and J. Zhang, ``An experimental study of the plasmonic Talbot effect", Opt. Express {\bf 17} 19757 (2009)

\bibitem{Talbot_SPP_nano} D. van Oosten, M. Spasenovi\'c, and L. Kuipers, ``Nanohole Chains for Directional and Localized Surface Plasmon Excitation", Nano Lett. {\bf 10}, 286 (2010).
 \bibitem{Berry96} M. V. Berry and S. Klein, ``Integer, fractional and fractal Talbot effects", J. Mod. Opt, {\bf 43}, 2139 (1996).
\bibitem{Berry} M. V. Berry, I. Marzoli, and W. P. Schleich, ``Quantum carpets, carpets of light", Phys. World  {\bf 14} 39 (2001).

\bibitem{Talbot_Rayleigh} Lord Rayleigh, ``On copying diffraction gratings, and some phenomena connected therewith" Phil. Magn. {\bf 11} 196 (1881)

\bibitem{Talbot_X ray imaing} F. Pfeiffer, T. Weitkamp, O. Bunk, and C. David, ``Phase retrieval and differential phase-contrast imaging with low-brilliance X-ray sources", Nat. Phys. {\bf 2}, 258 (2006).

\bibitem{BEC06} J. Kasprzak,  M. Richard, S. Kundermann, A. Baas, P. Jeambrun, J. M. J. Keeling, F. M. Marchetti, M. H. Szyman\'{s}ka, R. Andr\'e, J. L. Staehli, V. Savona, P. B. Littlewood, B. Deveaud, and Le Si Dang, ``Bose-Einstein condensation of exciton polaritons", Nature {\bf 443}, 409 (2006).
\bibitem{CiutiREV13} I. Carusotto and C. Ciuti, ``Quantum fluids of light", Rev. Mod. Phys. {\bf 85}, 299 (2013).
\bibitem{YamamotoREV14} T. Byrnes, N. Y. Kim, and Y. Yamamoto, ``Exciton-polariton condensatesÓ, Nat. Phys. {\bf 10}, 803 (2014).
\bibitem{stanford1d} C. W. Lai, N. Y. Kim, S. Utsunomiya, G. Roumpos, H. Deng, M. D. Fraser, T. Byrnes, P. Recher, N. Kumada, T. Fujisawa, and Y. Yamamoto, ``Coherent zero-state and $\pi$-state in an exciton-polariton condensate array", Nature {\bf 450}, 529 (2007).
 \bibitem{gap_state} D. Tanese, H. Flayac, D. Solnyshkov, A. Amo, A. Lema\^{\i}tre, E. Galopin, R. Braive, P. Senellart, I. Sagenes, G. Malpuech, and J. Bloch, ``Polariton condensation in solitonic gap states in a one-dimensional periodic potential", Nature Comm., {\bf 4}, 1749 (2013)
 \bibitem{acoustic_tight} E. A. Cerda-M\'endez, D. N. Krizhanovskii, K. Biermann, R. Hey, M. S. Skolnick, and P. V. Santos, ``Wavefunction of polariton condensates in a tunable acoustic lattice", New J. Phys., {\bf 14} 075011 (2012).
\bibitem{arXiv_15} K. Winkler, O. A. Egorov, I. G. Savenko, X. Ma, E. Estrecho, T. Gao, S. M\"{u}ller, M. Kamp, T. C. H. Liew, E. A. Ostrovskaya, S. H\"{o}fling, C. Schneider, ``Collective state transitions of exciton-polaritons loaded into a periodic potential", Phys. Rev. B {\bf 93}, 121303(R) (2016).
\bibitem{frustrated} F. Baboux, L. Ge, T. Jacqmin, M. Biondi, E. Galopin, A. Lema\^{\i}tre, L. Le Gratiet, I. Sagnes, S. Schmidt, H.?E. T\"ureci, A. Amo, and J. Bloch, Phys. Rev. Lett. {\bf 116}, 066402 (2016) 
 \bibitem{stanford2d} N. Y. Kim, K. Kusudo, C. Wu, N. Masumoto, A. L\"{o}ffler, S. H\"{o}fling, N. Kumada, L. Worschech, A. Forchel and Y. Yamamoto, ``Dynamical d-wave condensation of excitonÐpolaritons in a two-dimensional square-lattice potential", Nat. Phys. {\bf 7}, 681 (2011)
\bibitem{stanfordkagome} N. Masumoto, N. Y. Kim, T. Byrnes, K. Kusudo, S. H\"{o}fling, A. Forchel, and Y. Yamamotoâ ``Exciton-polariton condensates with flat bands in a two-dimensional kagome lattice", New J. Phys. {\bf 14}, 065002 (2012)

\bibitem{acoustic} E. A. Cerda-M\'endez, D. N. Krizhanovskii,  M. Wouters, R. Bradley, K. Biermann, K. Guda, R. Hey, P.V. Santos, D. Sarkar, and M. S. Skolnick, Phys. Rev. Lett. ``Exciton-polariton gap solitons in two-dimensional lattices {\bf 105}, 116402 (2010).

\bibitem{NJP_15} K. Winkler, J. Fischer, A. Schade, M. Amthor, R. Dall, J. Ge{\ss}ler, M. Emmerling, E. A. Ostrovskaya, M. Kamp, Ch. Schneider, and S. H\"{o}fling, ``A polariton condensate in a photonic crystal potential landscape", New J. Phys. {\bf 17}, 023001 (2015).
\bibitem{trapping_review} Ch. Schneider, K. Winkler, M.D. Fraser, M. Kamp, Y. Yamamoto, E. A. Ostrovskaya, S. H\"{o}fling, ``Exciton-Polariton Trapping and Potential Landscape Engineering", arXiv:1510.07540 (2015).
\bibitem{SM} Supplemental Material section contains details of polariton emission below condensation threshold, power dependence of the Talbot effect, and discussion of other types of Talbot carpets observable in this system.
\bibitem{lattice_PRL} E. A. Ostrovskaya, J. Abdullaev, M. D. Fraser, A. S. Desyatnikov, Y. S. Kivshar, ``Self-localization of polariton condensates in periodic potentials", Phys. Rev. Lett., {\bf 110}, 170407 (2013)
\bibitem{wires_2014} A. Mischok, V. G. Lyssenko, R. Br\"uckner, F. L\"ochner, R. Scholz, A. A. Zakhidov, H. Fr\"ob and K. Leo, ``Zero- and $\pi$-States in a Periodic Array of Deep Photonic Wires", Adv. Optical Mater. {\bf 2}, 746 (2014). 
\bibitem{wires_2015} A. Mischok, R. Br\"uckner, H. Fr\"ob, V. G. Lyssenko, K. Leo, and A. A. Zakhidov, ``Control of Lasing from Bloch States in Microcavity Photonic Wires via Selective Excitation and Gain", Phys. Rev. Applied {\bf 3}, 064016 (2015).

\bibitem{bec_gap} H. Pu, L.O. Baksmaty, W. Zhang, N.P. Bigelow, and P. Meystre, Phys. Rev. A {\bf 67}, 043605 (2003); B. Eiermann, Th. Anker, M. Albiez, M. Taglieber, P. Treutlein, K.-P. Marzlin, and M. K. Oberthaler, Phys. Rev. Lett. {\bf 92} 230401 (2004); P. J. Louis, E. A. Ostrovskaya,C. M. Savage, and Yu. S. Kivshar, Phys. Rev. A, {\bf 67}, 013602 (2003); N. K. Efremidis and D. N. Christodoulides, Phys. Rev. A, {\bf 67}, 063608 (2003); E. A. Ostrovskaya and Yu. S. Kivshar, Opt. Express {\bf 12}, 19 (2004).
\bibitem{Wouters2008} M. Wouters, I. Carusotto, and C. Ciuti, ``Spatial and spectral shape of inhomogeneous nonequilibrium exciton-polariton condensates", Phys. Rev. B {\bf 77}, 115340 (2008).
\bibitem{anharmonic}  T. J. Alexander, M. Salerno, E. A. Ostrovskaya, and Yu. S. Kivshar, ``Matter waves in anharmonic periodic potentials", Phys. Rev. A {\bf 77}, 043607 (2008).
\bibitem{Lohmann90} A. W. Lohmann and J. A. Thomas, ``Making an array illuminator based on the Talbot effect", Appl. Opt. {\bf 29} 4337 (1990).



\end{thebibliography}
\end{document}